\documentclass{eas}
\usepackage{graphicx}
\begin{document}

\TitreGlobal{Mass Profiles and Shapes of Cosmological Structures}

\title{A universal density slope -- velocity anisotropy relation}
\author{Hansen, S. H.}\address{University of Zurich, Winterthurerstrasse 190, 8057
Zurich, Switzerland}
\author{Moore, B.}
\author{Stadel, J.}
%
\runningtitle{A universal $\alpha-\beta$ relation}
\setcounter{page}{23}
\index{Author1, A.}
\index{Author2, B.}
\index{Author3, C.}

%
\begin{abstract} One can solve the Jeans equation analytically for
equilibrated dark matter structures, once given two pieces of input
from numerical simulations.  These inputs are 1) a connection between
phase-space density and radius, and 2) a connection between velocity
anisotropy and density slope, the $\alpha-\beta$ relation. The first
(phase-space density v.s. radius) has been analysed through several
different simulations, however the second ($\alpha-\beta$ relation) has not
been quantified yet. We perform a large set of numerical experiments
in order to quantify the slope and zero-point of the $\alpha-\beta$
relation. When combined with the assumption of phase-space being
a power-law in radius 
this allows us to conclude that equilibrated dark matter
structures indeed have zero central velocity anisotropy,
central density
slope of $\alpha_0 \approx -0.8$, and outer anisotropy of
approximately $\beta_\infty \approx 0.5$.  
\end{abstract}
\maketitle

%
\section{Introduction}

We have seen remarkable progress in the understanding of pure dark
matter structures over the last few years. This was triggered by
numerical simulations which have observed general trends in the
behaviour of the radial density profile of equilibrated dark matter
structures from cosmological simulations, which roughly follow an NFW
profile~\cite{nfw96,moore}, (see Diemand et al 2004 for references).
General trends in the radial dependence of the velocity anisotropy has
also been suggested~\cite{cole}.  Recently, more complex 
relations have been identified, holding even for systems that
do not follow the simplest radial power-law behaviour in density. These
relations are first that the phase-space density, $\rho/\sigma^3$, is a
power-law in radius~\cite{taylor}, and second that there is a linear
relationship between the density slope and the
anisotropy~(Hansen \& Moore 2005, hereafter HM04).
A connection between the shape of the velocity distribution
function and the density slope has also been 
suggested (Hansen et al 2005a, 2005b).

Using the Jeans equation together with the fact that phase-space
density is a power-law in radius allows one to find the
density slope in the central region numerically~\cite{taylor}, and even
analytically for power-law densities~\cite{jeanspaper}. Recently
Dehnen \& McLaughlin (2005, hereafter DM05)
using both the phase-space density being a power-law in radius and
also the $\alpha-\beta$ relation, showed
that one can solve the Jeans equations analytically and extract the
radial dependence of density, anisotropy, mass etc.

The relationship between phase-space density and radius has been
considered several times and
seems to be well established, however, the other crucial ingredient in
the analysis, namely the linear relationship between density slope and
anisotropy has only been investigated qualitatively.
We have performed a large set of simulations in order to quantify
this relationship. We show that with present day simulations
there does indeed appear to be a linear relation between density slope
and anisotropy, which implies zero anisotropy near the density slope
of approximately $-0.8$. We combine our findings with
analytical results and find that the outer anisotropy is 
radial and close to $+0.5$.

\section{The simulations}

The $\alpha-\beta$ relation was suggested based on a set of 
different numerical simulations, including spherical collapse
simulations, collisions between initially isotropic NFW structures,
merger between two disk galaxies, and finally structures formed
in a $\Lambda$CDM cosmological simulation. The original relation
had a large scatter, and there were several questions not fully 
answered,
such as a possible dependence on shape or initial conditions.
We have here performed controlled experiments to test exactly 
these points.
The set of experiments are described in glorious detail
in a coming publication, and we will not get into these aspects here.
We will only mention, that for each simulation have we
extracted all the relevant parameters in radial bins, 
logarithmically distributed from the softening length to beyond
the region which is fully equilibrated. We calculate the radial
derivative of the density (the density slope)
\begin{equation}
\alpha \equiv \frac{d {\rm ln} \rho}{d {\rm ln} r} \, ,
\end{equation}
and the velocity anisotropy
\begin{equation}
\beta \equiv 1 - \frac{\sigma^2_t}{\sigma^2_r} \, ,
\end{equation}
in each radial bin, where the $\sigma^2_{\rm t,r}$ are the one
dimensional velocity dispersions in the tangential and radial
directions respectively.


\subsection{Dependence on shape?}

One of the test we performed was the head-on collision between
two initially isotropic NFW structures. We then took the
resulting structures and collided these headon again.
The resulting structure is prolate when observed in
density contours. One may fear that the resulting $\alpha-\beta$ relation will
depend strongly on the axis-ratios of such structures. 
In order to test this question, we extracted the resulting prolate
structure after the first collision described above. We can now decide
to collide this structure along different axes, either along the long,
intermediate or short axes.  We perform these 3 different collisions.

We find that there is virtually no difference in the $\alpha-\beta$
relation for these 3 different simulations.
We conclude from
this that whereas the definition of density slope does depend
slightly on the shape of the density contours, then the $\alpha-\beta$ relation
is almost independent of the shape.

\begin{figure}[ht]
   \centering
   \includegraphics[width=9cm]{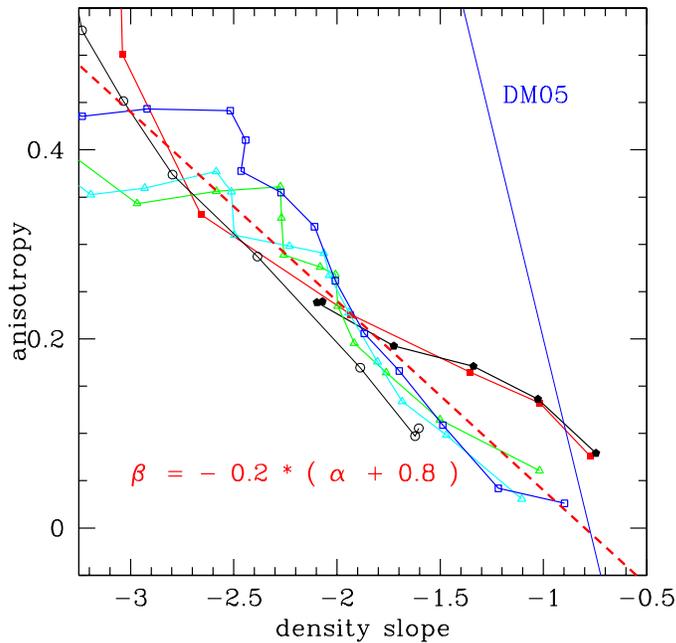}
      \caption{A collection of various simulations, 
including the tests of initial conditions
and shape. The red (dashed) line is a fit through the data.
The thin (blue, solid) line shows the theoretical {\em central} quantities
(HM05). The crossing of these two lines must be the true central values.}
       \label{fig:together2}
   \end{figure}

\subsection{Dependence on initial conditions?}

To address the question of how strongly the $\alpha-\beta$ relation from the
head-on collisions described above depend on the initial conditions,
we now perform collisions between structures with different 
initial degree of anisotropy. First we make a repeated head-on
collision between two isotropic NFW structures. Next we make
repeated head-on collisions between two initially strongly radially
anisotropic structures. And finally do we perform repeated
head-on collisions between initially strongly tangentially
anisotropic structures.

We find that the resulting $\alpha-\beta$ relation is virtually 
independent of initial conditions, and we conclude that
the collisions were sufficiently violent to erase the initial
conditions sufficiently to conform with the $\alpha-\beta$ relation.

\section{Comparison with theory}

In a recent paper, Dehnen \& McLaughlin (2005) showed, that under
the two assumption that phase-space density is a power-law 
in radius, and that there is a linear $\alpha-\beta$ relation, that
the central slope of dark matter structures must be
$\alpha_0 = -(7 + 10\beta_0)/9$,  where $\beta_0$ is the
central anisotropy. We can now compare this result with
our findings, and in figure~\ref{fig:together2} we show
the DM05 result for the {\em central} values as a thin (blue) 
straight line. We see that the two lines cross near $\beta=0$ and 
$\alpha=-7/9$, showing that the central part of dark matter structures
is isotropic, and that the central density slope is indeed $-7/9$.
The outer density slope is about $\gamma_\infty
= -31/9 \approx -3.44$ \cite{austin}, 
which when compared with our findings result in 
$\beta_\infty \approx 0.53$.  


\section{Conclusions}

We have quantified the relationship between the density slope and the
velocity anisotropy, the $\alpha-\beta$ relation. We have performed a
large set of simulations to investigate systematic effects related to
shape and initial conditions, and we find that the $\alpha-\beta$
relation is almost blind to both shape and initial conditions, as long
as the system has been {\em perturbed sufficiently} and subsequently
allowed to relax.  When compared with analytical results we find that
the central region is indeed isotropic with a density slope about
$0.8$, and that the outer asymptotic anisotropy is radial, with a
magnitude of $\beta \approx 0.5$.




\end{document}